\documentclass[12pt]{article}
\usepackage{putex}
\usepackage{graphicx}
\usepackage{enumerate}
\usepackage[nosort]{cite}
\usepackage{bbm}
\usepackage{amsmath}
\usepackage{amssymb}
\usepackage{amscd}
\usepackage{slashed}
\usepackage{array}
\usepackage{rotating}
\usepackage{float}

\newcommand{\grp}[1]{\mathrm{#1}}

\def\arccosh{\mop{arccosh}}

\newcommand{\abs}[1]{\lvert #1 \rvert}

\newcommand{\HGF}[4]{{}_2 F_1 \left(#1,#2;#3;#4\right)}
\newcommand{\calT}{{\cal T}_{\alpha\beta}^{(0)}}

\newcommand{\grU}{\grp{U}}
\newcommand{\grSU}{\grp{SU}}
\newcommand{\grSO}{\grp{SO}}
\newcommand{\superN}{\mathcal{N}}
\newcommand{\Integers}{\mathbbm{Z}}
\newcommand{\Reals}{\mathbbm{R}}

\newcommand{\AdS}{AdS}
\newcommand{\KK}{j}
\newcommand{\DeltaBPS}{\Delta_{\textrm{BPS}}}

\begin{document}

\preprint{PUPT-2296}

\institution{PU}{Joseph Henry Laboratories, Princeton University, Princeton, NJ 08544, USA}
\institution{PCTS}{Princeton Center for Theoretical Science, Princeton, NJ 08544, USA}

\title{The Squashed, Stretched, and Warped\\ Gets Perturbed}

\authors{Igor R. Klebanov,\worksat{\PU, \PCTS}\footnote{e-mail: {\tt klebanov@Princeton.EDU}} Silviu S. Pufu,\worksat{\PU}\footnote{e-mail: {\tt spufu@Princeton.EDU}} and Fabio D. Rocha\worksat{\PU}\footnote{e-mail: {\tt frocha@Princeton.EDU}}}

\abstract{We use direct Kaluza-Klein reduction to calculate the spectrum of spin-$2$ modes around a warped product of $AdS_4$ and a certain squashed and stretched 7-sphere. The modes turn out to be polynomials in the four complex variables parameterizing the sphere, and their complex conjugates.  The background, which possesses $\grU(1)_R\times\grSU(3)$ symmetry, has been conjectured to be dual to a $\grU(N)\times \grU(N)$ $\superN=2$ superconformal Chern-Simons theory with a sextic superpotential. We find that the $\grU(1)_R\times\grSU(3)$ quantum numbers of spin-$2$ modes are in agreement with those determined in arXiv:0809.3773 through a group theoretic method, and with the spectrum of spin-2 gauge invariant operators in the Chern-Simons gauge theory. The mass-squared in $AdS_4$ is found to be quadratic in these quantum numbers and the Kaluza-Klein excitation number. Most of the spin-$2$ operators belong to long multiplets, and we determine their dimensions via the AdS/CFT correspondence.}

\date{April 2009}

\maketitle

\tableofcontents

\section{Introduction}
\label{INTRODUCTION}

Superconformal Chern-Simons gauge theories are good candidates for describing the dynamics of coincident M2-branes \cite{Schwarz:2004yj}. Bagger and Lambert \cite{Bagger:2006sk, Bagger:2007jr, Bagger:2007vi}, and Gustavsson \cite{Gustavsson:2007vu} succeeded in constructing the first $\superN=8$ supersymmetric classical actions for Chern-Simons gauge fields coupled to matter.  Requiring manifest unitarity restricts the gauge group to $\grSO(4)$ \cite{Gauntlett:2008uf,Papadopoulos:2008sk}; this model may be reformulated as $\grSU(2)\times \grSU(2)$ gauge theory with conventional Chern-Simons terms having opposite levels $k$ and $-k$ \cite{VanRaamsdonk:2008ft,Bandres:2008vf}. For $k=2$ this model is believed to describe two M2-branes on the orbifold $\Reals^8/\Integers_2$ \cite{Lambert:2008et,Distler:2008mk}, but for other values of $k$ its interpretation is less clear. Aharony, Bergman, Jafferis, and Maldacena (ABJM) \cite{Aharony:2008ug} proposed that a similar $\grU(N)\times \grU(N)$ Chern-Simons gauge theory with levels $k$ and $-k$ arises on the world volume of $N$ M2-branes placed at the singularity of $\Reals^8/\Integers_k$, where $\Integers_k$ acts by simultaneous rotation in the four planes. Therefore, the ABJM theory was conjectured to be dual to M-theory on $AdS_4\times S^7/\Integers_k$. For $k>2$ this orbifold preserves only $\superN=6$ supersymmetry, and so does the ABJM theory \cite{Aharony:2008ug,Benna:2008zy,Bandres:2008ry}. The conjectured duality predicts that for $k=1,2$ the supersymmetry of the gauge theory must be enhanced to $\superN=8$. An interesting feature of the ABJM theory is the presence of certain ``monopole operators'' \cite{Hooft:1977hy,Borokhov:2002cg,Borokhov:2003yu} which create quantized flux of the diagonal $\grU(1)$ magnetic field. Their inclusion is expected to play a crucial role both in the enhancement of the supersymmetry and in describing the full spectrum of gauge invariant operators (see \cite{Berenstein:2008dc,Klebanov:2008vq,Imamura:2009ur,Kim:2009wb} for recent discussions of the monopole operators in this context).

As in the well understood examples of $AdS_5/CFT_4$ duality \cite{Maldacena:1997re,Gubser:1998bc,Witten:1998qj}, it is interesting to study Renormalization Group (RG) flows leading to super-conformal theories with lower supersymmetry. A well-known relevant superpotential deformation of the $\superN=4$ SYM theory by a term quadratic in one of the three chiral adjoint superfields, leads in the IR to an $\superN=1$ gauge theory with $\grU(1)_R \times \grSU(2)$ global symmetry and a quartic superpotential. This gauge theory, as well as the gravity dual of the RG flow, was studied in \cite{Freedman:1999gp}. Analogously, it is interesting to consider an ${\cal N} = 2$ superpotential deformation of the ABJM theory by a term quadratic in one of the four bi-fundamental superfields \cite{Benna:2008zy} (see also \cite{Ahn:2008ya}). This superpotential mass term requires the use of certain monopole operators that exist only at levels $k = 1,2$ \cite{Klebanov:2008vq}.  This relevant deformation leads to the IR theory with a sextic superpotential, which possesses $\grU(1)_R \times \grSU(3)$ global symmetry. Therefore, it was conjectured \cite{Benna:2008zy,Ahn:2008ya,Klebanov:2008vq} that this theory is dual to the $\grU(1)_R \times \grSU(3) $-invariant extremum \cite{Warner:1983vz} of the $\superN = 8$ gauged supergravity potential \cite{deWit:1982ig}.  In \cite{Corrado:2001nv}, this background was uplifted to a  warped product of $AdS_4$ and a ``squashed and stretched'' $7$-sphere.  Moreover, in \cite{Corrado:2001nv} the entire holographic RG flow was constructed from $AdS_4 \times S^7$ to this warped, squashed, and stretched background of M-theory. The relevant operator generating the RG flow was shown in \cite{Ahn:2000aq} to have dimension two, which agrees with the dimension of the fermion bilinear added to the action of ABJM theory in \cite{Benna:2008zy}.

One can further check this $AdS_4/CFT_3$ conjecture by comparing the $\grU(1)_R \times\grSU(3) $ quantum numbers and the dimensions of gauge-invariant operators in the IR ${\cal N} = 2$ superconformal Chern-Simons theory \cite{Benna:2008zy} with those of supergravity fluctuations around the background of \cite{Warner:1983vz,Corrado:2001nv}.  In \cite{Klebanov:2008vq}, the quantum numbers of all Kaluza-Klein (KK) supergravity excitations were computed using group theory methods introduced in \cite{Nicolai:1985hs}, without doing an explicit Kaluza-Klein reduction from $11$ to $4$ dimensions. However, this method does not determine the dimensions of operators that belong to long supermultiplets. The group theory alone gives two alternative ways of assigning $\grU(1)_R \times \grSU(3)$ quantum numbers consistent with ${\cal N} = 2$ SUSY \cite{Nicolai:1985hs}.  The first, called Scenario I in \cite{Klebanov:2008vq}, yields agreement with the gauge theory spectrum, while the second, Scenario II, is in disagreement with the gauge theory proposal.  The two scenarios give distinct mass spectra, so an explicit KK reduction would tell us which of the two scenarios is correct.

In this paper, we perform an explicit KK analysis of the spin-2 fields in $AdS_4$.  In the $11$-dimensional geometry, the equations describing these metric perturbations reduce to a minimally coupled scalar equation.  We find analytic solutions for all the KK modes; they turn out to be polynomials in the four complex variables parameterizing the squashed and stretched $\Reals^8$, and their complex conjugates.  Quite remarkably, the squared masses in $AdS_4$ for all these modes are quadratic functions of the $\grU(1)_R \times \grSU(3)$ quantum numbers, as well as of the KK excitation number.  These results hold not only for the BPS states, but also for the non-BPS ones. The spectrum that we find is indeed consistent with Scenario I, and therefore with the duality proposed in \cite{Benna:2008zy,Klebanov:2008vq}.

The rest of the paper is organized as follows.  In section~\ref{BACKGROUND} we review the $11$-dimensional background found in \cite{Corrado:2001nv} to be the uplifting of Warner's $\grU(1)_R \times \grSU(3)$ extremum of gauge supergravity.  In section~\ref{MINIMAL} we solve the minimally coupled scalar equation in this background and find its spectrum.  In section~\ref{MATCH} we describe the connection between the minimally coupled scalar equation and the $AdS_4$ graviton, and we match the quantum numbers of the operators that we find with those of operators in the Chern-Simons theory with sextic superpotential.  We end with a discussion of our results in section~\ref{DISCUSSION}.

\section{The background geometry}
\label{BACKGROUND}

We start by reviewing the $11$-dimensional uplift of the supergravity background with global $\grU(1)_R \times \grSU(3)$ symmetry that was found in \cite{Warner:1983vz} as a non-trivial extremum of the ${\cal N} =8$ gauged supergravity potential.  The $11$-dimensional geometry is a warped product of $AdS_4$ and an internal manifold, which in this case is a squashed and stretched $S^7$ \cite{Corrado:2001nv}.  This background has a non-zero four-form flux $F_{(4)} = dA_{(3)}$ in the $S^7$ directions.  As a result, parity is broken, which can also be seen from the fact that in the corresponding four-dimensional gauged supergravity background a scalar and a pseudo-scalar acquire VEVs.  We refer the reader to \cite{Corrado:2001nv} for a derivation of the formulae in this section.

In the conventions of \cite{Duff:1986hr}, the bosonic field equations are
\eqn{SUGRAeoms}{
R_{AB} + R g_{AB} = {1 \over 3} F_{AMNP}^{(4)} F_{B}^{(4)MNP}\,, \qquad d \ast F_{(4)} = F_{(4)} \wedge F_{(4)} \,,
}
where the Hodge dual is defined with the convention $\epsilon^{0\cdots 10} = 1$.  We focus first on the geometry of the $S^7$ part.  Following \cite{Corrado:2001nv}, we consider a space diffeomorphic to $\Reals^8$ parameterized by $x^I$, with $I$ running from $1$ to $8$, in which the $S^7$ will be embedded in the standard way.  In this 8-dimensional space, we introduce the K\"ahler form
\eqn{Jdef}{
J_{12} = J_{34}= J_{56} =J_{78} = 1\,,
}
and the diagonal matrix
\eqn{Qdef}{
Q = \textrm{diag}\left\{\rho^{-2},\rho^{-2},\rho^{-2},\rho^{-2},\rho^{-2},\rho^{-2}, \rho^6, \rho^6\right\} \,.
}
The metric on the deformed $\Reals^8$ is taken to be
\eqn{metricSSR8}{
ds^2_8(\rho,\chi) = g_{IJ} dx^I dx^J = dx^I Q^{-1}_{IJ} dx^J + {\sinh{\chi}^2 \over \xi^2} (x^I J_{IJ} dx^J)^2 \,,
}
where $\xi^2 \equiv x^I Q_{IJ} x^J$.  Equation \eqref{metricSSR8} describes a squashed and stretched $\Reals^8$; the amounts of squashing and stretching are parameterized by $\rho$ and $\chi$, respectively.  For $(\rho,\chi)=(1,0)$ we obtain the flat metric on $\Reals^8$, which has $SO(8)$ symmetry. For other values of $\rho$ and $\chi$, the $SO(8)$ symmetry group is broken down generically to $\grSU(3)\times \grU(1)^2$. This can be seen explicitly through introducing the complex coordinates
\eqn{zwDef}{
z^1 = x^1 + i x^2\,, \qquad z^2 = x^3 + i x^4\,, \qquad z^3 = x^5 + i x^6\,, \qquad w = x^7 -i x^8 \,.
}
It is straightforward to check that unitary rotations of the $z^i$, as well as the multiplication of $w$ by a phase, are isometries of the metric \eqref{metricSSR8}. This gives us the $\grU(3)\times \grU(1)= \grSU(3)\times \grU(1)^2$ isometry group.

We now introduce angular coordinates $y^\alpha \equiv (\mu, \theta, \alpha_1, \alpha_2, \alpha_3, \phi, \psi)$ parameterizing the $S^7$
$|z^1|^2+ |z^2|^2+ |z^3|^2+ |w|^2=1$
inside $\Reals^8$:
\eqn{S7inR8coords}{
z^1 & = \cos\mu \sin\theta \cos\left({\alpha_1\over 2}\right) e^{{i\over 2} (\alpha_2+\alpha_3)} e^{i(\phi+\psi)} \cr
z^2 & = \cos\mu \sin\theta \sin\left({\alpha_1\over 2}\right) e^{-{i\over 2} (\alpha_2-\alpha_3)} e^{i(\phi+\psi)} \cr
z^3 & = \cos\mu \cos \theta e^{i(\phi+\psi)} \cr
w &= \sin\mu\, e^{-i\psi} \,.
}
In order for the $y^\alpha$ to cover the $S^7$ only once, their ranges can be taken as follows:
\eqn[c]{yAlphaRanges}{
\begin{aligned}
0 &\leq \mu, \theta \leq \pi/2 &\qquad 0 &\leq \alpha_1 \leq \pi \\
0 &\leq \alpha_3, \phi, \psi \leq 2 \pi &\qquad -2 \pi &\leq \alpha_2 \leq 2 \pi \,.
\end{aligned}
}
Through the embedding \eqref{S7inR8coords}, the $7$-sphere inherits the stretching and squashing of the ambient space $\Reals^8$, so its metric is just the pullback of \eqref{metricSSR8}:
\eqn{metricSSS7}{
ds^2_7(\rho,\chi) = g_{\alpha\beta} dy^\alpha dy^b\,,
  \qquad g_{\alpha\beta} = {\partial x^I\over \partial y^\alpha}
  {\partial x^J\over \partial y^\beta} g_{IJ} \,.
}

The $11$-dimensional metric that solves \eqref{SUGRAeoms} (with an appropriately chosen $F_{(4)}$ given below) is then given by\footnote{In \eqref{metricSSS7}, $L$ is defined as $3^{-3/4}$ times the corresponding quantity appearing in \cite{Corrado:2001nv}.  We prefer this normalization because the radius of $AdS_4$ is now given by $L$ and not $3^{3/4} L$.}
\eqn{metric11d}{
ds^2_{11} = \Delta^{-1} ds^2_4 + 3^{3/2} L^2 \Delta^{1\over 2} ds^2_7 (\rho, \chi)\,,\qquad
\Delta \equiv (\xi \cosh \chi)^{-{4\over 3}} \,,
}
where $ds^2_4$ is the $AdS_4$ metric
\eqn{metricAdS4}{
ds^2_4 = e^{2r/L} \left[-(dx^0)^2 + (dx^1)^2 + (dx^2)^2 \right] + dr^2  \,,
}
and $\rho$ and $\chi$ are set to
\eqn{rhochivalues}{
\rho = 3^{1\over 8}\,, \qquad \chi = {1\over 2}\arccosh 2 \,.
}
Note that the warp factor $\Delta$ depends only on the angular coordinate $\mu$ and that it does not break any of the $\grSU(3)\times \grU(1)^2$ isometries of \eqref{metricSSS7}.

To describe the $F_{(4)}$ that solves \eqref{SUGRAeoms} together with \eqref{metric11d}, we start by constructing a $3$-form $C_{(3)}$ on the $S^7$ fiber.  In terms of the complex coordinates on $\Reals^8$ defined in \eqref{S7inR8coords}, $C_{(3)}$ can be written as
\eqn{GotC3}{
C_{(3)} = {3^{11/4} L^3 \over 4 \left(z^i \bar z_i + 3 w \bar w \right)}
  \left[z^{[1} dz^2 \wedge dz^{3]} \wedge d \bar w
  - \bar w dz^1 \wedge dz^2 \wedge dz^3 \right] \,.
}
One can then construct the $A_{(3)}$ in the $11$-dimensional geometry by taking\footnote{This doesn't fully agree with \cite{Corrado:2001nv}, but we checked that the Maxwell and Einstein equations are satisfied.}
\eqn{GotA3}{
A_{(3)} = {3^{3/4} \over 4} e^{3 r/L} dx^0 \wedge dx^1 \wedge dx^2
  + C_{(3)} + C_{(3)}^\ast \,.
}
The $4$-form $F_{(4)}$ appearing in \eqref{SUGRAeoms} is then $F_{(4)} = dA_{(3)}$.  The internal part of $F_{(4)}$  can be written as
\eqn{GotdC3}{
 d C_{(3)}  + d C_{(3)}^\ast &= {3^{11/4} L^3 \over 2 (1+2 w\bar w)^2} \bigg[-
  \bar w z^{[1} dz^2 \wedge dz^{3]} \wedge dw \wedge d\bar w
  - {\bar w}^2 dz^1 \wedge dz^2 \wedge dz^3 \wedge dw\cr
  {}&+ (1 + w \bar w) dz^1 \wedge dz^2 \wedge dz^3 \wedge d\bar w
  \bigg]  + \text{c.c.}
}

Note that $F_{(4)}$ breaks the $\grSU(3) \times \grU(1)^2$ symmetry group of \eqref{metric11d} to $\grSU(3) \times \grU(1)$.  In the complex coordinates \eqref{S7inR8coords} this symmetry group consists of $\grSU(3)$ rotations of the $z^i$, as well as transformations of the type
\eqn{RemainingU1}{
z^i \to z^i e^{i \delta}\,, \qquad w \to w e^{3 i \delta}
}
with arbitrary $\delta$.  In terms of the $y^\alpha$ coordinates, \eqref{RemainingU1} corresponds to shift symmetries of $\phi$ and $\psi$ that preserve the quantity $3 \phi + 4 \psi$.  This $\grU(1)$ should be identified with the R-symmetry of the dual field theory.  In order to agree with the convention used in \cite{Klebanov:2008vq}, we define the R-charge to be given by the Killing vector
\eqn{RChargeDef}{
R = -i\left({4 \over 3} \partial_\phi - \partial_\psi\right) 
  = {1 \over 3} \left(z^i \partial_{z^i} - \bar z_i \partial_{\bar z_i} \right)
   + w \partial_w - \bar w \partial_{\bar w} \,.
}
Thus, the $z^i$ coordinates have R-charge $1/3$ \cite{Johnson:2001ze}, while the $w$ coordinate has R-charge $1$.

\section{Minimally coupled scalar equation}
\label{MINIMAL}

The action for a minimally coupled scalar in the background described in the previous section is given by
\eqn{Sscalar}{
S= \int d^{11}x\, \sqrt{-g} \left[-  {1\over 2} \left( \partial\phi\right)^2 \right] \,,
}
The equation of motion following from this action is
\eqn{ScalarEOM}{
\Box \phi = 0 \,,
}
where $\Box$ denotes the $11$-dimensional laplacian.  Making the separation of variables ansatz
\eqn{SepOfVars}{
\phi=\Phi(x^i,r) Y(y^\alpha) \,,
}
we can write \eqref{ScalarEOM} as
\eqn{scalareom11d}{
Y(y^\alpha)  \Box_{4} \Phi(x^i,r) +  \Phi(x^i,r)  {\cal L}  Y(y^\alpha) =0 \,,
}
in which $\Box_4$ denotes the $AdS_4$ laplacian and ${\cal L}$ is a differential operator acting on the squashed and stretched $S^7$ given by
\eqn{s7opdef}{
{\cal L} \equiv {\Delta^{-1} \over \sqrt {-g_{11}}} \partial_{\alpha} \left( \sqrt{-g_{11}} g^{\alpha\beta}_{11}\partial_\beta\right) = {\Delta^{-3/4} \over \sqrt{g_7}} \partial_{\alpha} \left(\Delta^{-3/4} \sqrt{g_7} g^{\alpha\beta}_{7} \partial_\beta \right) \,.
}
Here, $g^{7}_{\alpha\beta}$ and $g^{11}_{\mu\nu}$ are the metrics \eqref{metricSSS7} and \eqref{metric11d}, respectively.

If we now choose $Y(y^\alpha)$ to be an eigenfunction of this differential operator, namely
\eqn{EvalueProblem}{
{\cal L} Y(y^\alpha)= -m^2 Y(y^\alpha) \,,
}
then \eqref{scalareom11d} becomes the equation of motion of a massive scalar field in $AdS_4$,
\eqn{AdS4MassiveScalar}{
\Box_4 \Phi(r,x^i) - m^2 \Phi(r,x^i) = 0 \,.
}
The 11-dimensional minimally coupled scalar thus gives a tower of 4-dimensional Kaluza-Klein modes that are all massive $AdS_4$ scalars with masses determined by the eigenvalues of ${\cal L}$.

To compute these eigenvalues we should exploit the symmetries of the metric. Let us first consider the $\grSU(3)$ piece of the isometry group. In the $(z^i,w)$ coordinates defined in \eqref{S7inR8coords}, the Killing vectors associated to the $\grSU(3)$ symmetry are simply given by
\eqn{KillingVectorsInz}{
{\boldsymbol \xi} = T^k_{\phantom{k}l} z^l \partial_{z^k}  - (T^k_{\phantom{k}l})^* \bar z_l \partial_{\bar z_k}, \qquad k,l=1,2,3 \,,
}
where $T^k_{\phantom{k}l}$ are arbitrary traceless hermitian matrices.  A convenient set of linearly independent Killing vectors is obtained by taking $T=\lambda^a/2$, where $\lambda^a$ with $a=1,\ldots,8$ are the Gell-Mann matrices.  In an irreducible representation of $\grSU(3)$ labeled by the Dynkin labels $[p, q]$, the quadratic Casimir
\eqn{QuadCasimir}{
{\cal C}_2 \equiv \sum_{a = 1}^8 \boldsymbol{\xi}^a \boldsymbol{\xi}^a
}
has eigenvalues
\eqn{C2EValues}{
{\cal C}_2(p, q) = {1\over 3}q^2 + {1\over 3} p^2 + {1\over 3} p q + p+q \,.
}
In the holomorphic coordinates $(z^i, w)$ its eigenfunctions are, up to normalization, just linear combinations of products between $p$ factors of $z^i$ and $q$ factors of $\bar z_i$, namely
\eqn{C2Efunctions}{
Y_{pq} (z, \bar z)=
  a_{i_1 i_2 \ldots i_p}^{j_1 j_2 \ldots j_q}
  \left( \prod_{k = 1}^p z^{i_k} \right) \left( \prod_{l = 1}^q \bar z_{j_l} \right)\,.
}
Here, $a_{i_1 i_2 \ldots i_p}^{j_1 j_2 \ldots j_q}$ is a $(p, q)$-tensor independent of the $(z^i, w)$ coordinates that is symmetric in its lower indices, symmetric in its upper indices, and satisfies the tracelessness condition $a_{k i_2 \ldots i_p}^{k j_2 \ldots j_q} = 0$.
Since ${\cal C}_2$ doesn't act on $\mu$ or on $\psi$, we can multiply the expression \eqref{C2Efunctions} by any function of $w$ and $\bar w$.

We will look for eigenmodes of the form\footnote{One might think that this ansatz is not general enough since \eqref{YinzwAgain} can be multiplied by $e^{i n_\phi \phi}$ with some integer $n_\phi$.  However, one can check that if $n_\phi \neq 0$ this extra factor either makes the wavefunction ill-defined at $\theta = \pi/2$ or turns it into a wavefunction of the form \eqref{YinzwAgain} that can be obtained from the $[p + n_\phi, q - n_\phi]$ representation of $\grSU(3)$.}
\eqn{YinzwAgain}{
Y(y^\alpha) = Y_{pq}(z, \bar z)
   w^{n_r} H(u) \,, \qquad u \equiv 1- w\bar w = \cos^2 \mu\ .
}
The R-charge of this wave-function \eqref{YinzwAgain} can be read off from the powers of $z^i$, $\bar z_i$, and $w$ that appear in this formula if one recalls that \eqref{RChargeDef} implies that the $z^i$ coordinates carry R-charge $1/3$, the $\bar z_i$ coordinates carry R-charge $-1/3$, and that the $ w$ coordinate carries R-charge $1$.  Therefore,
\eqn{GotRForEigenFns}{
R = {1\over 3} (p - q) + n_r \,.
}

Using the definition of the quadratic Casimir \eqref{QuadCasimir} together with the formulae for $\boldsymbol{\xi}^a$ given in \eqref{KillingVectorsInz}, one can show that the eigenvalue problem \eqref{EvalueProblem} reduces to
\eqn{YtildeEq}{
  (1-u)u H'' + \left(c - (a_-+a_++1) u \right) H' -a_-a_+ H =0 \,,
}
where primes denote derivatives with respect to $u$, and
\eqn{acVals}{
a_\pm &= \frac{1}{6} \bigg[ 9+3 p+3 q + 3 n_r \pm  \bigg( 81 + 18 m^2 L^2 + 7(p^2+q^2) + 10 p q \cr &\qquad {}- 6(p-q)n_r+24(p+q) - 9 n_r^2 \bigg)^\frac{1}{2} \bigg] \cr
c &= 3+p+q \,.
}
This equation is solved by hypergeometric functions.  The boundary conditions that enforce regularity of $Y(y^\alpha)$ are that $u^{{p + q \over 2}} H(u)$ is regular at $u = 0$ and that $(1 - u)^{{n_r \over 2}} H(u)$ is regular at $u = 1$.  The solution for which $u^{{p + q \over 2}} H(u)$ is regular at $u = 0$ is
\eqn{Ysol}{
H(u) = \HGF{a_-}{a_+}{c}{u} \,.
}
In order to impose the boundary conditions at $u = 1$, we start by noting that when $a_-=-\KK$, with $\KK$ a non-negative integer, the hypergeometric function $ \HGF{a_-}{a_+}{c}{u}$ reduces to a polynomial in $u$ of order $j$. If $n_r\geq 0$, $(1 - u)^{n_r \over 2} H(u)$ is therefore regular at $u = 1$. For $n_r<0$, if we set $a_-=n_r-\KK$, with $\KK$ still a non-negative integer, then $H(u)$ has a zero of order $n_r$ at $u=1$ and $(1-u)^{n_r\over 2}  H(u)$ is once again well-behaved. It is straightforward but tedious to check that if $a_-$ does not have one of these forms, equation \eqref{YtildeEq} doesn't have solutions that lead to well-behaved $(1-u)^{n_r\over 2} H(u)$.  The KK spectrum of the minimally coupled scalar is then obtained by setting $a_-$ as given by \eqref{acVals} equal to $-\KK$ if $n_r\geq 0$ and to $n_r-\KK$ if $n_r<0$, and solving for $m^2$. The result can be written compactly in terms of $\KK$, $n_r$, $p$, and $q$ as
\eqn{GotLambda}{
m^2 &= {1 \over L^2} \bigg[2\KK^2
  + 2 \KK \abs{n_r} + n_r^2 + 2 \KK (p + q + 3)
  + {1\over 3} n_r \left(p - q\right) \cr
  {}&+ \abs{n_r} (3 + p + q) + {1\over 9} \left(p^2 + q^2 + 4 p q + 15 p
  + 15 q \right) \bigg] \,.
}
It is interesting that $m^2$ is given by such a simple quadratic formula.

Plugging \eqref{Ysol} into \eqref{YinzwAgain} and using the appropriate formulae for $a_\pm$ and $c$ we see that
\eqn{EFunctionsInz}{
Y(y^\alpha) &= a_{i_1 i_2 \ldots i_p}^{j_1 j_2 \ldots j_q}
  \left(\prod_{k = 1}^p z^{i_k} \right) \left(\prod_{l = 1}^q \bar z_{j_l} \right) w^{n_r} \cr
 {}&\times \begin{cases}
    {}_2F_1(-\KK, 3 + p + q + \KK + n_r; 3 + p + q; 1 - w \bar w) & \text{if $n_r\geq 0$}\\
    {}_2F_1(-\KK + n_r, 3 + p + q + \KK; 3 + p + q; 1 - w \bar w) & \text{if $n_r<0$} \,.
 \end{cases}
}
As we remarked, the hypergeometric functions appearing in \eqref{EFunctionsInz} are in fact polynomials in their last argument. For example,
\eqn{Examples}{
\KK = 0, n_r \geq 0 &: \qquad Y(y^\alpha) \sim Y_{pq}(z,\bar z) w^{n_r} \cr
\KK = 1, n_r \geq 0 &: \qquad Y(y^\alpha) \sim Y_{pq} (z, \bar z)
  w^{n_r} {-(n_r + 1) + (4 + n_r + p + q) w \bar{w} \over 3 + p + q} \,.
}
To obtain the eigenfunctions when $n_r<0$, one just needs to interchange $w$ and $\bar{w}$ in \eqref{Examples}.

\section{Spin-2 Kaluza-Klein spectrum}
\label{MATCH}

The spectrum of the minimally coupled scalar obtained in the previous section is in fact the same as that of a  graviton polarized in the $AdS_4$ directions.  Such gravitons correspond to fluctuations of the metric \eqref{metric11d},
\eqn{MetricFluctuations}{
 g_{mn} \to g_{mn} + h_{mn}
}
with $m$ and $n$ referring to the $AdS_4$ coordinates $y^m = (x^0, x^1, x^2, r)$.  We can choose a gauge where $h_{rm} = 0$, the only remaining non-zero metric fluctuations being $h_{ij}$, with $i$ and $j$ running from $0$ to $2$.  In addition to this, we require
\eqn{hijConditions}{
h^i_{\phantom{i}i} = 0\,, \qquad \partial^i h_{ij} = 0 \,.
}
The conditions \eqref{hijConditions} can be thought of as projecting out the spin-0 and spin-1 components of the graviton multiplet.  As in the unwarped AdS metrics, the linearized Einstein equations reduce to the minimally coupled scalar equation for $\phi = h^i_{\phantom{i}j}$ \cite{Gubser:1997yh, Constable:1999gb}.  It immediately follows that the KK spectrum of the spin-2 component of the graviton multiplet is the same as that of a massless scalar.

The dimensions of the CFT operators dual to the KK modes \eqref{EFunctionsInz} can be calculated from the standard AdS/CFT relation
\eqn{GotDelta}{
\Delta (\Delta-3) = m^2 L^2 \,,
}
where $m^2$ is given in \eqref{GotLambda}.  Note that \eqref{GotDelta} applies both to the scalar and the spin-2 modes (see, for example, \cite{D'Hoker:2002aw}).  The R-charge of these operators is given by \eqref{GotRForEigenFns}.  Recall that $p$ and $q$ appearing in \eqref{GotRForEigenFns}--\eqref{EFunctionsInz} are the Dynkin labels of $\grSU(3)$ irreducible representations $[p, q]$; $n_r$ is an arbitrary integer, which according to \eqref{GotRForEigenFns}, is in one-to-one correspondence with the R-charge for fixed $p$ and $q$; and $\KK$ is a non-negative integer, the Kaluza-Klein excitation number.

We would like to compare the spectrum of the minimally coupled scalar to the two scenarios in  \cite{Klebanov:2008vq}.  The only ${\rm Osp}(2|4)$ supermultiplets with spin-2 components are the massless, short, and long graviton multiplets, denoted in \cite{Klebanov:2008vq} by MGRAV, SGRAV, and LGRAV, respectively.  For both MGRAV and SGRAV, the supersymmetry shortening conditions require $\Delta = |R| + 3$, giving $m^2 = m_{\rm BPS}^2$ with
\eqn{mBPSDef}{
m_{\rm BPS}^2 \equiv {1\over L^2} \abs{R} \left(\abs{R} + 3 \right) \,.
}
The dimensions of the spin-2 operators in LGRAV satisfy $\Delta > \DeltaBPS $, and consequently $m^2 > m_{\rm BPS}^2$.

Let's denote multiplets belonging to the $[p, q]$ representation of $\grSU(3)$ which have R-charge $R$ by $[p, q]_R$.   In both scenarios in \cite{Klebanov:2008vq} there is a unique massless graviton multiplet whose quantum numbers are $[0, 0]_0$.  In Scenario I, all the short graviton multiplets are $[0, 0]_{r}$ and $[0, 0]_{-r}$, with $r$ a positive integer.  In Scenario II, there is an infinite number of short graviton multiplets of the form $[0, 0]_0$, as well as short graviton multiplets with non-zero $p$ and $q$ such as $[1, 0]_{-2/3}$, $[0, 1]_{2/3}$, $[2, 0]_{4/3}$, $[0, 2]_{-4/3}$, etc.  Plugging \eqref{GotRForEigenFns} into \eqref{mBPSDef} and comparing to \eqref{GotLambda}, one can check that $m^2 \geq m_{\rm BPS}^2$ with equality only when $p = q = \KK = 0$.  The equality case corresponds exactly to operators belonging to $[0, 0]_R$ with all integer $R$. This is in agreement with Scenario I and in disagreement with Scenario II. The structure of the long graviton multiplets predicted by the R-charge formula \eqref{GotRForEigenFns} is also in agreement with Scenario I and in disagreement with Scenario II; this can be seen by examining Tables~17 through~23 in \cite{Klebanov:2008vq}.

We can go further and identify the operators dual to the modes described by \eqref{EFunctionsInz}.   Let us discuss these operators schematically, as in \cite{Klebanov:2008vq}, in terms of bifundamental matter superfields ${\cal Z}^A$ with $A$ ranging from $1$ to $4$, as well as gauge superfields. We will not be careful with gauge indices, and assume that appropriate insertions of monopole operators make the resulting expressions gauge invariant. The gauge theory conjectured to be dual to the $\grU(1)_R \times \grSU(3)$ ${\cal N} = 2$ supergravity background examined in this paper is a deformation of ABJM theory by a superpotential term quadratic in ${\cal Z}^4$.  The gauge theory also has $\grU(1)_R \times \grSU(3)$ symmetry, where the $\grSU(3)$ symmetry corresponds to global rotations of ${\cal Z}^1$, ${\cal Z}^2$, and ${\cal Z}^3$ into one another.  Under the $\grU(1)_R$ symmetry, the fields ${\cal Z}^A$ have R-charges given by
\eqn{calZRcharges}{
R({\cal Z}^1) = R({\cal Z}^2) = R({\cal Z}^3) = {1 \over 3}\,,  \qquad R({\cal Z}^4) = 1 \,.
}
In \cite{Klebanov:2008vq}, it was proposed that the gauge theory operators dual to the short graviton multiplets $[0, 0]_r$ with $n\geq 0$ are of the schematic form
\eqn{calTShort}
{
{\cal T}^{(n)}_{\alpha\beta} \sim \calT ({\cal Z}^4)^r \,,
}
where $\calT$ is the stress-energy superfield
\eqn{calTDef}{
\calT = \bar D_{(\alpha} \bar{\cal Z}_A D_{\beta)}  {\cal Z}^A
  + i \bar{\cal Z}_A {\overleftrightarrow \partial}_{\alpha\beta} {\cal Z}^A \,.
}
The operator $\calT$ is dual to the massless graviton multiplet $[0, 0]_0$ and is conserved.  The form \eqref{calTShort} is only schematic because in order to properly define gauge invariant operators of this form one needs to include Dirac monopole operators:  see \cite{Klebanov:2008vq}.  As mentioned above, short multiplets have $p = q = \KK = 0$.  From \eqref{GotRForEigenFns}, \eqref{EFunctionsInz}, and \eqref{Examples} we can see that the spin-$2$ components of these multiplets have
\eqn{WavefunctionShort}{
Y(y^\alpha) = w^r \,,
}
for R-charge $r\geq 0$.

\begin{table}[t!]
\caption{The first few spin-$2$ components of the graviton multiplets.  For each multiplet, we give the Dynkin labels $[p, q]$, the R-charge $R$, the values of $\KK$ and $n_r$, the dimension $\Delta$ of the spin-$2$ component of the multiplet, the mass $m^2 L^2$ of the dual $AdS_4$ field, and a schematic form of the dual CFT operator.  The dimension $\Delta$ can be computed from $m^2 L^2$ as the larger root of equation \eqref{GotDelta}.  The operators marked with ``$*$'' are BPS.}
\label{Modes}
\begin{center}
\begin{tabular}{rl|c|c|c|c|c}
& $[p, q]_R$ & $\KK$ & $n_r$ & $\Delta$ & $m^2 L^2$ & Operator \\
\hline \hline
* & $[0, 0]_0$ & $0$ & $0$ & $3$ & $0$ & $\calT$ \\
* & $[0, 0]_{\pm 1} $ & $0$ & $\pm 1$ & $4$ & $4$
   & $\calT {\cal Z}^4$, $\calT \bar{{\cal Z}}_4$ \\
& $[0, 1]_{-{1\over 3}}$, $[1, 0]_{1 \over 3}$ & $0$
   & $0$ & ${1 \over 6} \left(9 + \sqrt{145} \right)$
   & ${16 \over 9}$ & $\calT \bar{{\cal Z}}_A$, $\calT {\cal Z}^A$ \\
* & $[0, 0]_{\pm 2}$ & $0$ & $\pm 2$ & $5$ & $10$
   & $\calT ({\cal Z}^4)^2$, $\calT (\bar{{\cal Z}}_4)^2$ \\
& $[0, 0]_{0}$ & $1$ & $0$ & ${1 \over 2} \left(3 + \sqrt{41} \right)$
   & ${8}$ & $\calT \left(1 - 4a^2 {\cal Z}^4 \bar{\cal Z}_4 \right)$ \\
& $[0, 1]_{-{4 \over 3}}$, $[1, 0]_{4 \over 3}$ & $0$ & $-1$, $1$
   & ${1 \over 6} \left(9 + \sqrt{337} \right)$ & ${64 \over 9}$
   & $\calT \bar{\cal Z}_A \bar{\cal Z}_4$, $\calT {\cal Z}_A {\cal Z}^4$ \\
& $[0, 1]_{{2 \over 3}}$, $[1, 0]_{-{2 \over 3}}$ & $0$ & $-1$, $1$
   & ${1 \over 6} \left(9 + \sqrt{313} \right)$ & ${58 \over 9}$
   & $\calT \bar{\cal Z}_A {\cal Z}^4$, $\calT {\cal Z}_A \bar{\cal Z}_4$ \\
& $[0, 2]_{-{2 \over 3}}$, $[2, 0]_{{2 \over 3}}$ & $0$ & $0$
   & ${1 \over 6} \left(9 + \sqrt{217} \right)$ & ${34 \over 9}$
   & $\calT \bar{\cal Z}_{(A} \bar{\cal Z}_{B)}$, $\calT {\cal Z}^{(A} {\cal Z}^{B)}$ \\
& $[1, 1]_{0}$ & $0$ & $0$ & $4$ & $4$
   & $\calT \left( {\cal Z}^{A} \bar{\cal Z}_{B} - {1 \over 3} \delta^A_B {\cal Z}^C \bar{\cal Z}_C \right)$\\
& $[0, 0]_{\pm 1}$ & $1$ & $\pm 1$ & ${1\over 2}(3 + \sqrt{65})$ & $14$
   & $\calT  \left( 2  - 5 a^2 {\cal Z}^4 \bar{\cal Z}_4 \right) {\cal Z}^4,
   \textrm{c.c.}$\\
* & $[0, 0]_{\pm 3}$ & $0$ & $\pm 3$ & $ 6 $ & $18$
   & $\calT \left( {\cal Z}^4 \right)^3,\calT \left( \bar{\cal Z}_4 \right)^3$\\
& $[1, 0]_{-{5\over 3}}, [0,1]_{5\over 3} $ & $0$ & $-2,+2$ & $ {1\over 6} (9 + \sqrt{553}) $ & $118\over 9$
   & $\calT {\cal Z}^A \left( \bar{\cal Z}_4 \right)^2,\calT \bar{\cal Z}_A \left( {\cal Z}^4 \right)^2 $\\
& $[1, 0]_{1\over 3}, [0,1]_{-{1 \over 3}} $ & $1$ & $0$ & $ {1\over 6} (9 + \sqrt{505}) $ & $106\over 9$
   & $\calT {\cal Z}^A \left(1 - 5 a^2 \bar{\cal Z}_4{\cal Z}^4  \right),\textrm{c.c.}$\\
& $[1, 0]_{7\over 3}, [0,1]_{-{7 \over 3}} $ & $0$ & $2,-2$ & $ {1\over 6} (9 + \sqrt{601}) $ & $130\over 9$
  & $\calT {\cal Z}^A \left({\cal Z}^4\right)^2,\calT \bar{\cal Z}_A \left(\bar{\cal Z}_4\right)^2  $\\
& $[1, 1]_{\pm1}$ & $0$ & $\pm 1$ & $ 5 $ & $10$
   & $\calT  \left( {\cal Z}^{A} \bar{\cal Z}_{B} - {1 \over 3} \delta^A_B {\cal Z}^C \bar{\cal Z}_C \right) {\cal Z}^4, \textrm{c.c.}$\\
& $[2, 0]_{-{1\over 3}},[0, 2]_{{1\over 3}}$ & $0$ & $-1,1$ & $ {1\over 6}(9+\sqrt{409}) $ & $82\over 9$
   &$\calT{\cal Z}^{(A} {\cal Z}^{B)}  \bar{\cal Z}_4, \calT  \bar{\cal Z}_{(A} \bar{\cal Z}_{B)} {\cal Z}^4 $\\
& $[2, 0]_{{5\over 3}},[0, 2]_{-{5\over 3}}$ & $0$ & $1,-1$ & $ {1\over 6}(9+\sqrt{457}) $ & $94\over 9$
   &$\calT {\cal Z}^{(A} {\cal Z}^{B)}{\cal Z}^4 , \calT\bar{\cal Z}_{(A} \bar{\cal Z}_{B)}  \bar{\cal Z}_4 $\\
& $[2, 1]_{1\over 3},[1, 2]_{-{1\over 3}}$ & $0$ & $0$ & $ {1\over 6}(9+\sqrt{313}) $ & $58\over 9$
   &$\calT \left( {\cal Z}^{(A}{\cal Z}^{B)} \bar{\cal Z}_{C} - {1 \over 3} \delta^{(A}_C {\cal Z}^{B)}_{\phantom{C}} {\cal Z}^D \bar{\cal Z}_D \right),\textrm{c.c.}$\\
& $[3, 0]_{1}, [0,3]_{-1}$ & $0$ & $0$ & $ {1\over 2} (3+\sqrt{33}) $ & $6$
   & $\calT {\cal Z}^{(A} {\cal Z}^B {\cal Z}^{C)}, \calT \bar{\cal Z}_{(A} \bar{\cal Z}_B \bar{\cal Z}_{C)}$
\end{tabular}
\end{center}
\end{table}%

It is then natural to identify (up to normalization) the ${\cal Z}^A$ fields, where $A = 1, 2, 3$, with the holomorphic coordinates $z^i$, and ${\cal Z}^4$ with $ w$.  From \eqref{EFunctionsInz}, one can then read off the operators corresponding to each of the KK modes.  In Table~\ref{Modes} we list a few of these modes, and we give a schematic form of the dual gauge theory operators.

\section{Discussion}
\label{DISCUSSION}

In this paper we performed a KK reduction for spin-$2$ excitations around a warped M-theory background which was conjectured in \cite{Benna:2008zy, Klebanov:2008vq} to be dual to an ${\cal N} = 2$ deformation of ABJM theory with $\grU(1)_R \times \grSU(3)$ symmetry.  This background is a warped product between $AdS_4$ and a squashed and stretched $S^7$ \cite{Corrado:2001nv}. The spectrum of spin-$2$ excitations was found by solving the equations of motion for a minimally coupled scalar in this background.  Our main results are equations \eqref{GotLambda} and \eqref{EFunctionsInz} that give the $AdS_4$ masses of the KK modes and their wavefunctions on the internal manifold.  It is remarkable that the squared masses of these modes are given by a simple quadratic function of all the quantum numbers, namely the Dynkin labels $[p, q]$ of $\grSU(3)$ representations, a $\grU(1)$ excitation number $n_r$ related to the R-charge through \eqref{GotRForEigenFns}, and the KK excitation number $j$.   In \cite{Klebanov:2008vq}, group theory methods were used to constrain the spectrum of supergravity fluctuations of the same background. The spectrum that we found agrees with Scenario~I and rules out Scenario~II, in agreement with the proposal of \cite{Klebanov:2008vq}.  Using the AdS/CFT duality, we computed the dimensions of the dual operators in the boundary CFT\@.   We proposed a schematic form of these operators in Table~\ref{Modes}.

An intriguing feature of the spectrum we obtained is the presence of modes with integer dimension $\Delta$ that do not belong to BPS multiplets.  For instance, the modes with quantum numbers $[1,1]_r$ and  $j=0$, with $r$ an integer, have R-charge $R = r$ and dimension $\Delta=|R|+4$, which is strictly greater than the corresponding BPS value $|R| + 3$.  The first three modes in this tower, $[1,1]_0$ and $[1,1]_{\pm 1}$, appear in Table~\ref{Modes}. The dual gauge theory operators corresponding to the spin-$2$ components of $[1, 1]_r$ are of the schematic form
\eqn{LongIntTowerposr}{
\begin{aligned}
 &T_{\alpha\beta} \left( { Z}^{A} \bar{ Z}_{B} - {1 \over 3} \delta^A_B { Z}^C \bar{ Z}_C \right) \left( Z^4\right)^{r}  &\qquad &\text{for $r\geq 0$} \\
 &T_{\alpha\beta} \left( { Z}^{A} \bar{ Z}_{B} - {1 \over 3} \delta^A_B { Z}^C \bar{ Z}_C \right) \left( \bar{Z}^4\right)^{-r} &\qquad &\text{for $r < 0$} \,,
\end{aligned}
}
where $Z^A$ are the spin-$0$ components of the ${\cal Z}^A$ superfields we used in the previous section.  In the rest of this discussion let's focus on the $r \geq 0$ case, the $r<0$ case being entirely analogous.   We recognize that the operators in \eqref{LongIntTowerposr} are products of two BPS protected operators:  the spin-$2$ component of the short graviton multiplets $[0, 0]_r$ (see \eqref{calTShort})
\eqn{ShortGravitonOp}{
T_{\alpha\beta}^{(r)}=T_{\alpha\beta} (Z^4)^{r}
}
with dimension $\Delta = |R| + 3$, and a scalar operator
\eqn{ScalarFromJ}{
{ Z}^{A} \bar{ Z}_{B} - {1 \over 3} \delta^A_B { Z}^C \bar{ Z}_C
}
with dimension $\Delta = 1$ belonging to the massless vector multiplet.  (Recall from \cite{Klebanov:2008vq} that the massless vector multiplet is dual to a conserved vector superfield ${\cal J}^{(0)B}_A$ whose spin-$1$ component is
\eqn{JConserved}{
J^{(0)A}_{\mu B} = \bar{Z}_B \overleftrightarrow{\partial}_\mu Z^A - \frac{1}{3} \delta^A_B \bar{Z}_C \overleftrightarrow{\partial}_\mu Z^C \,.
}
Being a conserved current, $J^{(0)A}_{\mu B}$ has protected dimension $\Delta=2$.)  The dimensions $\Delta= |R|+4$ of the operators \eqref{LongIntTowerposr} can therefore be correctly computed by naively adding the dimensions of the BPS operators \eqref{ShortGravitonOp} and \eqref{ScalarFromJ}.  We do not know of any mechanisms that protect the dimensions of the operators \eqref{LongIntTowerposr}.  It is worth noting that the $[1,1]_r$ modes are not the only ones producing integers dimensions:  there are infinitely many other such towers of long multiplets.  For instance, the modes $[3,6]_r$ and $[6,3]_{-r}$ with integer $r\geq -1$ and $j = 0$ have $\Delta=|R|+8$; as another example, the modes $[4,10]_r$ and $[10,4]_r$, with integer $r\geq -2$ and $j=0$ have dimensions $\Delta=|R| + 11$.  In addition, there are many other long multiplets with integer dimensions that do not belong to any such towers.

The existence of many towers of long multiplets with rational dimensions in the KK spectrum has been noted for other M-theory and string theory backgrounds.  This feature was pointed out for $\AdS_5\times T^{1,1}$ in \cite{Gubser:1998vd,Ceresole:1999rq}, for both $\AdS_4\times Q^{1,1,1}$ and $\AdS_4\times M^{1,1,1}$ in \cite{Fabbri:1999hw}, for $\AdS_4\times V_{(5,2)}$ in \cite{Ceresole:1999zg}, and for $\AdS_4\times N^{0,1,0}$ in \cite{Billo:2000zr}. For  $\AdS_4\times Q^{1,1,1}$ and $\AdS_4\times M^{1,1,1}$ the operators dual to the long rational gravitons are products of the stress tensor, a conserved current, and a chiral operator \cite{Fabbri:1999hw,Fabbri:1999ag}. More generally, it was shown in \cite{Billo:2000zs} that the KK spectrum of all M-theory backgrounds of the form $\AdS_4\times X_7$, where $X_7$ is a homogeneous space with Killing spinors, includes long multiplets with rational dimensions that appear as ``shadows'' of BPS-protected multiplets.  In the terminology of \cite{Billo:2000zs}, long graviton multiplets of a form analogous to \eqref{LongIntTowerposr} are shadows of short vector multiplets.  The reason behind the shadowing mechanism is that the same harmonics on $X_7$ appear in the KK expansion of two or more fields belonging to different multiplets.  The $AdS_4$ masses of these fields are related algebraically because they can each be expressed in terms of the eigenvalues of the same $X_7$ harmonics. There is no known interpretation of the shadowing mechanism in dual gauge theory language; nor is it known whether it survives the departure from the strong coupling limit, corresponding to including the string sigma model corrections.

It would be interesting to extend the analysis done in this paper to KK excitations of different $AdS_4$ spin. This  would permit further checks of Scenario~I of \cite{Klebanov:2008vq} and would perhaps elucidate the form of the gauge theory operators dual to these lower-spin excitations.  Our analysis was made easier by the fact that there was only one spin-$2$ excitation (given by certain perturbations of the metric with both indices in the $AdS_4$ directions) that decoupled from all other perturbations.  For lower spins, there are several distinct excitations corresponding to each spin, and one faces the additional challenge of finding the form of the perturbations that decouple.  This is made harder by the relatively small amount of symmetry in this background, and by the fairly involved expressions for the background metric and $3$-form.

\section*{Acknowledgments}
We thank T.~Klose and A.~Murugan for many useful discussions.  The work of I.R.K. was supported by the NSF grant number PHY-0756966.  S.S.P. and F.D.R. were supported in part by the NSF under award number PHY-0652782.  F.D.R. was also supported in part by the FCT grant SFRH/BD/30374/2006.

\clearpage
\bibliographystyle{ssg}
\bibliography{ssw}

\begingroup\raggedright\begin{thebibliography}{10}

\bibitem{Schwarz:2004yj}
J.~H. Schwarz, ``{Superconformal Chern-Simons theories},'' {\em JHEP} {\bf 11}
  (2004) 078, \href{http://xxx.lanl.gov/abs/hep-th/0411077}{{\tt
  hep-th/0411077}}.

\bibitem{Bagger:2006sk}
J.~Bagger and N.~Lambert, ``{Modeling multiple M2's},'' {\em Phys. Rev.} {\bf
  D75} (2007) 045020, \href{http://xxx.lanl.gov/abs/hep-th/0611108}{{\tt
  hep-th/0611108}}.

\bibitem{Bagger:2007jr}
J.~Bagger and N.~Lambert, ``{Gauge Symmetry and Supersymmetry of Multiple
  M2-Branes},'' {\em Phys. Rev.} {\bf D77} (2008) 065008,
  \href{http://xxx.lanl.gov/abs/0711.0955}{{\tt 0711.0955}}.

\bibitem{Bagger:2007vi}
J.~Bagger and N.~Lambert, ``{Comments On Multiple M2-branes},'' {\em JHEP} {\bf
  02} (2008) 105, \href{http://xxx.lanl.gov/abs/0712.3738}{{\tt 0712.3738}}.

\bibitem{Gustavsson:2007vu}
A.~Gustavsson, ``{Algebraic structures on parallel M2-branes},''
  \href{http://xxx.lanl.gov/abs/0709.1260}{{\tt 0709.1260}}.

\bibitem{Gauntlett:2008uf}
J.~P. Gauntlett and J.~B. Gutowski, ``{Constraining Maximally Supersymmetric
  Membrane Actions},'' \href{http://xxx.lanl.gov/abs/0804.3078}{{\tt
  0804.3078}}.

\bibitem{Papadopoulos:2008sk}
G.~Papadopoulos, ``{M2-branes, 3-Lie Algebras and Plucker relations},'' {\em
  JHEP} {\bf 05} (2008) 054, \href{http://xxx.lanl.gov/abs/0804.2662}{{\tt
  0804.2662}}.

\bibitem{VanRaamsdonk:2008ft}
M.~van Raamsdonk, ``{Comments on the Bagger-Lambert theory and multiple M2-
  branes},'' \href{http://xxx.lanl.gov/abs/0803.3803}{{\tt 0803.3803}}.

\bibitem{Bandres:2008vf}
M.~A. Bandres, A.~E. Lipstein, and J.~H. Schwarz, ``{N = 8 Superconformal
  Chern--Simons Theories},'' {\em JHEP} {\bf 05} (2008) 025,
  \href{http://xxx.lanl.gov/abs/0803.3242}{{\tt 0803.3242}}.

\bibitem{Lambert:2008et}
N.~Lambert and D.~Tong, ``{Membranes on an Orbifold},'' {\em Phys. Rev. Lett.}
  {\bf 101} (2008) 041602, \href{http://xxx.lanl.gov/abs/0804.1114}{{\tt
  0804.1114}}.

\bibitem{Distler:2008mk}
J.~Distler, S.~Mukhi, C.~Papageorgakis, and M.~Van~Raamsdonk, ``{M2-branes on
  M-folds},'' {\em JHEP} {\bf 05} (2008) 038,
  \href{http://xxx.lanl.gov/abs/0804.1256}{{\tt 0804.1256}}.

\bibitem{Aharony:2008ug}
O.~Aharony, O.~Bergman, D.~L. Jafferis, and J.~Maldacena, ``{N=6 superconformal
  Chern-Simons-matter theories, M2-branes and their gravity duals},'' {\em
  JHEP} {\bf 10} (2008) 091, \href{http://xxx.lanl.gov/abs/0806.1218}{{\tt
  0806.1218}}.

\bibitem{Benna:2008zy}
M.~Benna, I.~Klebanov, T.~Klose, and M.~Smedback, ``{Superconformal
  Chern-Simons Theories and $AdS_4/CFT_3$ Correspondence},'' {\em JHEP} {\bf
  09} (2008) 072, \href{http://xxx.lanl.gov/abs/0806.1519}{{\tt 0806.1519}}.

\bibitem{Bandres:2008ry}
M.~A. Bandres, A.~E. Lipstein, and J.~H. Schwarz, ``{Studies of the ABJM Theory
  in a Formulation with Manifest SU(4) R-Symmetry},'' {\em JHEP} {\bf 09}
  (2008) 027, \href{http://xxx.lanl.gov/abs/0807.0880}{{\tt 0807.0880}}.

\bibitem{Hooft:1977hy}
G.~'t~Hooft, ``{On the Phase Transition Towards Permanent Quark Confinement},''
  {\em Nucl. Phys.} {\bf B138} (1978) 1.

\bibitem{Borokhov:2002cg}
V.~Borokhov, A.~Kapustin, and X.-k. Wu, ``{Monopole operators and mirror
  symmetry in three dimensions},'' {\em JHEP} {\bf 12} (2002) 044,
  \href{http://xxx.lanl.gov/abs/hep-th/0207074}{{\tt hep-th/0207074}}.

\bibitem{Borokhov:2003yu}
V.~Borokhov, ``{Monopole operators in three-dimensional N = 4 SYM and mirror
  symmetry},'' {\em JHEP} {\bf 03} (2004) 008,
  \href{http://xxx.lanl.gov/abs/hep-th/0310254}{{\tt hep-th/0310254}}.

\bibitem{Berenstein:2008dc}
D.~Berenstein and D.~Trancanelli, ``{Three-dimensional N=6 SCFT's and their
  membrane dynamics},'' \href{http://xxx.lanl.gov/abs/0808.2503}{{\tt
  0808.2503}}.

\bibitem{Klebanov:2008vq}
I.~Klebanov, T.~Klose, and A.~Murugan, ``{$AdS_4/CFT_3$ -- Squashed, Stretched
  and Warped},'' {\em JHEP} {\bf 03} (2009) 140,
  \href{http://xxx.lanl.gov/abs/0809.3773}{{\tt 0809.3773}}.

\bibitem{Imamura:2009ur}
Y.~Imamura, ``{Monopole operators in N=4 Chern-Simons theories and wrapped
  M2-branes},'' \href{http://xxx.lanl.gov/abs/0902.4173}{{\tt 0902.4173}}.

\bibitem{Kim:2009wb}
S.~Kim, ``{The complete superconformal index for N=6 Chern-Simons theory},''
  \href{http://xxx.lanl.gov/abs/0903.4172}{{\tt 0903.4172}}.

\bibitem{Maldacena:1997re}
J.~M. Maldacena, ``{The large N limit of superconformal field theories and
  supergravity},'' {\em Adv. Theor. Math. Phys.} {\bf 2} (1998) 231--252,
  \href{http://xxx.lanl.gov/abs/hep-th/9711200}{{\tt hep-th/9711200}}.

\bibitem{Gubser:1998bc}
S.~S. Gubser, I.~R. Klebanov, and A.~M. Polyakov, ``{Gauge theory correlators
  from non-critical string theory},'' {\em Phys. Lett.} {\bf B428} (1998)
  105--114, \href{http://xxx.lanl.gov/abs/hep-th/9802109}{{\tt
  hep-th/9802109}}.

\bibitem{Witten:1998qj}
E.~Witten, ``Anti-de Sitter space and holography,'' {\em Adv. Theor. Math.
  Phys.} {\bf 2} (1998) 253--291,
  \href{http://xxx.lanl.gov/abs/hep-th/9802150}{{\tt hep-th/9802150}}.

\bibitem{Freedman:1999gp}
D.~Z. Freedman, S.~S. Gubser, K.~Pilch, and N.~P. Warner, ``{Renormalization
  group flows from holography supersymmetry and a c-theorem},'' {\em Adv.
  Theor. Math. Phys.} {\bf 3} (1999) 363--417,
  \href{http://xxx.lanl.gov/abs/hep-th/9904017}{{\tt hep-th/9904017}}.

\bibitem{Ahn:2008ya}
C.~Ahn, ``{Holographic Supergravity Dual to Three Dimensional N=2 Gauge
  Theory},'' {\em JHEP} {\bf 08} (2008) 083,
  \href{http://xxx.lanl.gov/abs/0806.1420}{{\tt 0806.1420}}.

\bibitem{Warner:1983vz}
N.~P. Warner, ``{Some new extrema of the scalar potential of gauged N=8
  supergravity},'' {\em Phys. Lett.} {\bf B128} (1983) 169.

\bibitem{deWit:1982ig}
B.~de~Wit and H.~Nicolai, ``{N=8 Supergravity},'' {\em Nucl. Phys.} {\bf B208}
  (1982) 323.

\bibitem{Corrado:2001nv}
R.~Corrado, K.~Pilch, and N.~P. Warner, ``{An N = 2 supersymmetric membrane
  flow},'' {\em Nucl. Phys.} {\bf B629} (2002) 74--96,
  \href{http://xxx.lanl.gov/abs/hep-th/0107220}{{\tt hep-th/0107220}}.

\bibitem{Ahn:2000aq}
C.-h. Ahn and J.~Paeng, ``{Three-dimensional SCFTs, supersymmetric domain wall
  and renormalization group flow},'' {\em Nucl. Phys.} {\bf B595} (2001)
  119--137, \href{http://xxx.lanl.gov/abs/hep-th/0008065}{{\tt
  hep-th/0008065}}.

\bibitem{Nicolai:1985hs}
H.~Nicolai and N.~P. Warner, ``{The $SU(3) \times U(1)$ invariant breaking of
  gauged N=8 supergravity},'' {\em Nucl. Phys.} {\bf B259} (1985) 412.

\bibitem{Duff:1986hr}
M.~J. Duff, B.~E.~W. Nilsson, and C.~N. Pope, ``{Kaluza-Klein Supergravity},''
  {\em Phys. Rept.} {\bf 130} (1986) 1--142.

\bibitem{Johnson:2001ze}
C.~V. Johnson, K.~J. Lovis, and D.~C. Page, ``{The Kaehler structure of
  supersymmetric holographic RG flows},'' {\em JHEP} {\bf 10} (2001) 014,
  \href{http://xxx.lanl.gov/abs/hep-th/0107261}{{\tt hep-th/0107261}}.

\bibitem{Gubser:1997yh}
S.~S. Gubser, I.~R. Klebanov, and A.~A. Tseytlin, ``{String theory and
  classical absorption by three-branes},'' {\em Nucl. Phys.} {\bf B499} (1997)
  217--240, \href{http://xxx.lanl.gov/abs/hep-th/9703040}{{\tt
  hep-th/9703040}}.

\bibitem{Constable:1999gb}
N.~R. Constable and R.~C. Myers, ``{Spin-two glueballs, positive energy
  theorems and the AdS/CFT correspondence},'' {\em JHEP} {\bf 10} (1999) 037,
  \href{http://xxx.lanl.gov/abs/hep-th/9908175}{{\tt hep-th/9908175}}.

\bibitem{D'Hoker:2002aw}
E.~D'Hoker and D.~Z. Freedman, ``{Supersymmetric gauge theories and the AdS/CFT
  correspondence},'' \href{http://xxx.lanl.gov/abs/hep-th/0201253}{{\tt
  hep-th/0201253}}.

\bibitem{Gubser:1998vd}
S.~S. Gubser, ``{Einstein manifolds and conformal field theories},'' {\em Phys.
  Rev.} {\bf D59} (1999) 025006,
  \href{http://xxx.lanl.gov/abs/hep-th/9807164}{{\tt hep-th/9807164}}.

\bibitem{Ceresole:1999rq}
A.~Ceresole, G.~Dall'Agata, R.~D'Auria, and S.~Ferrara, ``{Superconformal field
  theories from IIB spectroscopy on AdS(5) x T(11)},'' {\em Class. Quant.
  Grav.} {\bf 17} (2000) 1017--1025,
  \href{http://xxx.lanl.gov/abs/hep-th/9910066}{{\tt hep-th/9910066}}.

\bibitem{Fabbri:1999hw}
D.~Fabbri {\em et.~al.}, ``{3D superconformal theories from Sasakian
  seven-manifolds: New nontrivial evidences for AdS(4)/CFT(3)},'' {\em Nucl.
  Phys.} {\bf B577} (2000) 547--608,
  \href{http://xxx.lanl.gov/abs/hep-th/9907219}{{\tt hep-th/9907219}}.

\bibitem{Ceresole:1999zg}
A.~Ceresole, G.~Dall'Agata, R.~D'Auria, and S.~Ferrara, ``{M-theory on the
  Stiefel manifold and 3d conformal field theories},'' {\em JHEP} {\bf 03}
  (2000) 011, \href{http://xxx.lanl.gov/abs/hep-th/9912107}{{\tt
  hep-th/9912107}}.

\bibitem{Billo:2000zr}
M.~Billo, D.~Fabbri, P.~Fre, P.~Merlatti, and A.~Zaffaroni, ``{Rings of short N
  = 3 superfields in three dimensions and M-theory on AdS(4) x N(0,1,0)},''
  {\em Class. Quant. Grav.} {\bf 18} (2001) 1269--1290,
  \href{http://xxx.lanl.gov/abs/hep-th/0005219}{{\tt hep-th/0005219}}.

\bibitem{Fabbri:1999ag}
D.~Fabbri, ``{Three dimensional conformal field theories from Sasakian
  seven-manifolds},'' \href{http://xxx.lanl.gov/abs/hep-th/0002255}{{\tt
  hep-th/0002255}}.

\bibitem{Billo:2000zs}
M.~Billo, D.~Fabbri, P.~Fre, P.~Merlatti, and A.~Zaffaroni, ``{Shadow
  multiplets in AdS(4)/CFT(3) and the super-Higgs mechanism},'' {\em Nucl.
  Phys.} {\bf B591} (2000) 139--194,
  \href{http://xxx.lanl.gov/abs/hep-th/0005220}{{\tt hep-th/0005220}}.

\end{thebibliography}\endgroup
\end{document}